# Electrical injection and detection of spin-polarized currents in topological insulator $Bi_2Te_2Se$

Jifa Tian[1,2,*], Ireneusz Miotkowski[1], Seokmin Hong[2,3], and Yong P. Chen[1,2,3,*]

1. Department of Physics and Astronomy, Purdue University, West Lafayette, IN 47907, USA

2. Birck Nanotechnology Center, Purdue University, West Lafayette, IN 47907, USA

3. School of Electrical and Computer Engineering, Purdue University, West Lafayette, IN 47907, USA

*Email: tian5@purdue.edu; yongchen@purdue.edu

Abstract

Topological insulators (TIs) are an unusual phase of quantum matter with nontrivial spin-momentum locked topological surface states (TSS). The electrical detection of spin-momentum locking of the TSS in 3D TIs has been lacking till very recently. Many of the results are measured on samples with significant bulk conduction, such as metallic $Bi_2Se_3$, where it can be challenging to separate the surface and bulk contribution to the measured spin signal. Here, we report spin potentiometric measurements in thin flakes exfoliated from bulk insulating 3D TI $Bi_2Te_2Se$ (BTS221) crystals, using two outside nonmagnetic (Au) contacts for driving a DC spin helical current and a middle ferromagnetic (FM)-$Al_2O_3$ tunneling contact for detecting spin polarization. The voltage measured by the FM electrode exhibits a hysteretic step-like change when sweeping an in-plane magnetic field between opposite directions along the easy axis of the FM contact to switch its magnetization. Importantly, the direction of this step-like voltage change can be reversed by reversing the direction of the DC current, and the amplitude of the change as measured by the difference in the detector voltage between opposite FM magnetization increases linearly with increasing bias current, consistent with the current-induced spin polarization of spin-momentum-locked TSS. Our work directly demonstrates the electrical injection and detection of spin polarization in TI and may enable utilization of spin-helical TSS for future applications in nanoelectronics and spintronics.

Three-dimensional (3D) topological insulators (TIs) represent an interesting new class of quantum matter hosting spin helical topological surface states (TSS, in the TI bulk band gap, sketched in Fig. 1a) protected by time-reversal symmetry (TRS).[1-7] Many exotic physics (eg. majorana fermions and topological magnetoelectric effect [3,4,8]) have been predicted based on TIs and TSS. One of the most fundamental and striking properties of TSS is the spin-momentum locking (Fig. 1a inset, schematically showing spins tangential to the TSS Fermi surface and perpendicularly locked to the momentum). Such spin-momentum locking is the basis of the topological protection, where a backscattering that reverses momentum would reverse the spin thus is suppressed without breaking the TRS. It also gives rise to spin-helical current, where a directional electrical current ($\vec{I}$) carried by the TSS would be automatically spin-polarized, with spin polarization $S$ along the direction $\vec{n} \times \vec{I}$, where $\vec{n}$ is the surface normal. The spin polarization direction is in-plane and perpendicular to the current, and would reverse upon reversing the current direction or going to the opposite surface (reversing $\vec{n}$). Note the spin polarization direction on a given surface and for a given current direction is the same regardless whether the current is carried by electrons or holes (corresponding to Fermi level above and below the Dirac point, with opposite spin-*momentum* helicities [9]). The spin-helical current makes TI particularly promising for spintronic device applications, eg., using such current induced spin polarization for all-electric spin injection, etc. [4,10-15].

The helical spin texture of the spin-momentum-locked TSS in 3D TIs has been well established by spin and angle-resolved photoemission spectroscopy (spin ARPES) [9,16-22]. Further evidence has also been reported in experiments studying spin-polarized photocurrent [23], spin transfer torque [24], and spin-pumping [25] in TIs. The direct all-electrical detection of spin helical current or current induced spin polarization from spin-momentum-locked TSS in 3D TIs using spin-sensitive transport measurements is a key step for potential applications in spintronics. However, such all-electrical measurements have been challenging in TIs. Only very recently, a few experiments [26-31] have used ferromagnetic (FM) electrodes to detect current-induced spin polarization in MBE-grown $Bi_2Se_3$ thin films by four-terminal spin potentiometric measurements [26,27,28], exfoliated $Bi_2Se_3$ thin flakes by two-terminal spin valve measurements [29], MBE-grown $(Bi_{0.53}Sb_{0.47})_2Te_3$ thin films with DC+AC measurements [30], and exfoliated $Bi_{1.5}Sb_{0.5}Te_{1.7}Se_{1.3}$ flakes by non-local measurements [31]. Commonly studied TIs such as $Bi_2Se_3$ often have significant metallic bulk conduction, making it more challenging to measure and utilize the spin-polarized transport from TSS in the presence of parallel conducting channels from the bulk or the bulk band bending induced Rashba two dimensional electron gas (2DEG, which even possesses opposite spin polarization from the TSS [32,33]). Thus experiments on bulk-insulating TIs are highly desired to confirm and better access and utilize the current-induced surface spin polarization from TSS. $Bi_2Te_2Se$ (BTS221) is one of the first and best-studied bulk insulating TIs to allow better access of the TSS transport [34,35]. However, it has not been explored in spin transport experiments to measure the current-induced spin polarization from TSS.

In this report, we have fabricated 3D TI-based spin devices from thin flakes exfoliated from bulk BTS221 crystals. We performed 3-terminal spin potentiometric measurements [36], using two outside nonmagnetic contacts to inject a DC current and a middle ferromagnetic (FM) contact to detect the current-induced spin polarization on the top surface. The voltage between the FM and one of the nonmagnetic contacts is monitored as a function of an in-plane magnetic field applied to magnetize the FM contact along the easy axis (perpendicular to the current). We observe a hysteretic step-like voltage change when sweeping the magnetic field between opposite directions, resulting in a clear difference (asymmetry) in the voltage detected between opposite FM magnetizations. The "polarity" of this step-like voltage change reverses when the direction of the DC current is reversed. The amplitude of this voltage change (voltage difference between opposite FM magnetizations) is found to increase with increasing DC current but decrease with increasing temperature, and offers a clear electrical signal for the current-induced spin polarization from the TSS detected by our spin potentiometric measurement.

**Fabrication of TI spin devices and the spin potentiometric measurement.** Our high quality bulk-insulating 3D TI single crystals (BTS221) are grown by the Bridgeman method [37-39]. Thin flakes (typical thickness ~10 nm- 60 nm) are exfoliated from the bulk crystals using the "scotch-tape method" [40] and placed on top of heavily doped Si substrates (coated with 300 nm $SiO_2$). To fabricate the thin tunneling barrier (used for spin detection) and protect the sample surface from oxidation, the exfoliated flakes are immediately transferred into a high vacuum e-beam evaporator and coated with a 0.7 nm-thick Al film under a pressure of $2.0 \times 10^{-8}$ Torr. After re-exposing the sample in air, the Al film is fully oxidized into $Al_2O_3$, which has been widely used as a tunneling barrier in spin transport measurements. The FM (50 nm-thick permalloy (Py), $Ni_{0.80}Fe_{0.20}$) and non-magnetic (70-nm thick Au) contacts are fabricated by two rounds of e-beam lithography and e-beam evaporation. Prior to depositing the Au contacts, the samples are dipped in a BOE ($HF:H_2O=1:6$) solution to remove the $Al_2O_3$ in the contact area, so that Ohmic contacts can be achieved. Standard 4-terminal resistance measured as a function of temperature in reference devices fabricated using flakes exfoliated from the same BTS221 crystal and only Au contacts (without $Al_2O_3$ layer) shows the signatures of an insulating bulk with surface conduction at low temperature (Fig. S1). Low temperature field effect and Hall measurements (not shown) on the reference devices show the carriers are n-type.

The spin potentiometric measurement is performed in a three-terminal configuration involving one ferromagnetic (FM) "inner" contact (Py) and two non-magnetic (NM) "outer" contacts (Au), schematically shown in Fig. 1b. A DC source-drain bias current is applied between the two Au electrodes and a voltage (potential difference) is measured between the Py and Au (left) contacts using a high-impedance voltmeter. We define a positive DC bias current as flowing from right to left along –x direction, and a positive in-plane magnetic field as applied along the easy axis of the Py electrode in the -y direction. All the spin potentiometric measurements are performed under a DC bias current $I$ and an in-plane magnetic field in a variable temperature cryostat with the base temperature of 1.6 K.

**Theoretical understanding of the spin potentiometric measurement in TI.** We first discuss the theoretical principles behind the spin potentiometric measurements (working in the linear response regime) to probe the current induced spin polarization arising from the TSS on the top surface (Fig. 1c). The discussions below also assume the Fermi level ($E_f$) is located above the Dirac point (consistent with the n-type conduction in our devices, while similar discussions and conclusions also apply for the p-type case). Under a positive current ($I > 0$, with the corresponding electron current $i_e$ or momentum $k_e$ along the +x direction, Fig. 1d upper), the right-going ($k_e$ along +x) electrons have higher occupation and electrochemical potential (also known as "quasi Fermi-level") than those of the left-going ($k_e$ along -x) electrons [41]. Since the right (left)-going electrons have down (up) spins due to the spin-momentum locking of TSS (now down or "↓" spin means a spin polarized along –y direction) , this results in a spin electrochemical potential difference ($\mu_\downarrow - \mu_\uparrow$, proportional to $I$) to build up in the channel, giving rise to a non-equilibrium accumulation of down spins on the top surface or a current-induced surface spin polarization in the -y direction, as shown in the schematic surface state band structure (Fig. 1d, note for a 2D TSS the momentum $k_e$ and spin directions in the simplified schematics here are understood as the dominant (or averaged) momentum and spin polarization directions under a DC bias current).[36] In the potentiometric measurement, a FM detector (assumed to only weakly couple to the channel to probe the local potential without drawing significant current) with magnetization (which can be tuned by an in-plane magnetic field along the easy axis (y) of the FM electrode) along either -y or +y direction (defined as positive ("+M") or negative ("-M") magnetization, corresponding to the *majority spin* direction along +y or -y direction, respectively) is assumed to mainly measure the local $\mu_\uparrow$ or $\mu_\downarrow$ (Ref. 42, see also supplemental discussions in Fig. S2), making the voltage measured between the FM and the left Au to be $V_{+M}=(\mu_\uparrow - \mu_L)/(-e)$ or $V_{-M}=(\mu_\downarrow - \mu_L)/(-e)$ respectively (the minus sign in front of e reflects the negative electron charge). Since $\mu_\downarrow > \mu_\uparrow$, we have $V_{+M} > V_{-M}$. As the magnetic field (*B*) sweeps between positive and negative values (passing the corresponding coercive field of the FM to magnetize it along the *B* field direction), we expect the voltage signal measured by the FM detector to exhibit a hysteretic step-like change between the higher $V_{+M}$ and lower $V_{-M}$, as schematically depicted in the bottom panel of Fig. 1d. Importantly, reversing the bias current (*I* < 0) will make $\mu_\downarrow < \mu_\uparrow$ and reverse the TSS spin-polarization in the TI channel, therefore we expect the direction of the hysteretic step-like change in the measured voltage to also reverse, as now $V_{+M} < V_{-M}$ (see Fig. 1e). In summary, one can expect a low voltage or high potential when *M* (magnetization) of the FM detector is antiparallel to the channel spin polarization *S* (equivalently, when the FM majority spins are parallel to *S*, or when the magnetization of the channel is parallel to *M*, see Fig. S2), while a high voltage or low potential when *M* is parallel to *S* (the FM majority spins antiparallel to *S*, equivalently the magnetization of the channel antiparallel to *M*). This difference in detector voltage between opposite magnetizations, $\delta V = V_{+M} - V_{-M}$, will be the spin signal (of the channel spin polarization) measured in our experiment.

**Electrical detection of the current-induced TSS spin polarization in BTS221.** Figure 2 shows the key result of our TI spin potentiometric measurements, performed under a DC bias current (*I* = ±10 µA) at T=1.6 K in device *A*, which is fabricated from an exfoliated 40 nm-thick BTS221 flake. Applying an in-plane magnetic field (B, along y direction) orthogonal to the current direction (x direction), the Py detector magnetization *M* can be switched to be either parallel or antiparallel to the current-induced spin polarization *s* in the TI surface. The voltage (V) measured by the Py detector shows a clear hysteretic step-like change when sweeping the magnetic field, as shown in Fig. 2. Characteristic relative orientations of current *I* , *s* of TSS and *M* of detector (Py) are shown in the insets. As seen in Fig. 2a, under large positive magnetic fields and a positive current *I* =10 µA, *M* starts as parallel to *s*, and a higher voltage is measured and largely persists even when the magnetic field is decreased to zero (black trace). As the magnetic field sweeps to negative values passing the coercive field (~-7 mT) of the Py contact (to reverse its *M* to be antiparallel to the TSS spin polarization *s*), an abrupt decrease in the detector voltage is observed, exhibiting a step-like change (from higher to lower values). The trace features a pronounced *asymmetry* or difference (δV labeled by the arrow in Fig. 2a) in the voltage signal between large positive and negative magnetic fields. Reversing the magnetic field sweep from negative to positive values (red trace), the Py detector *M* switches at the positive coercive field of ~+7 mT (giving the hysteresis compared to the black trace), where the voltage abruptly increases as *M* now becomes again parallel to the spin orientation *s*. Most importantly, when the current is reversed (*I* =-10 µA), the orientation of the TI spin polarization *s* is also reversed and the trend of voltage change flips (δV reverses), as shown in Fig. 2b (there is a small, ~20µV, instrument-related DC offset in the voltage V. However, such an offset is independent of the current direction as well as magnetic field, and does not enter our measured spin signal δV). Our observation is qualitatively consistent with the expectations discussed in Figs. 1c-e. The sign of the current induced spin polarization measured in our sample is consistent with the spin helicity of the TSS at top surface [43], demonstrating the electrical detection of the spin-momentum locking of TSS in our samples (note the bottom surface carries the opposite spin polarization, which however is not detected by the FM contact on the top surface). We also note that a local sharp peak (eg. around ~-7mT in the down sweep, black curve, Fig. 2a) is sometimes observed in the detector voltage at the coercive field as the Py magnetization reverses. Such a feature may be attributed to the non-uniform reversal of magnetic domains [26] (that also leads to anomalous magnetic magnetoresistance (AMR) [31]) in the Py.

**Current and temperature dependences of the spin signal in BTS221.** We further studied the current dependence of the spin signal, measured by the voltage change or asymmetry discussed above, in device *A* at different bias currents (*I*) ranging from ±0.5 µA to ±40 µA. The representative results are shown in Fig. 3. We can see again the voltage asymmetry (step-like change) between large positive and negative magnetic fields (with a clear hysteresis near zero field) at bias currents as low as ±0.5 µA (Figs. 3a,b). We also note that both the overall scale of the voltage and the amplitude of the voltage change increase as *I* increases from ±0.5 µA to ± 40 µA (Figs. 3i,j), as more electrons with polarized spins from the TSS flow through the channel. The amplitude of the voltage change δV that detects the channel spin polarization is quantitatively extracted as δV=$V_{+M}$ -$V_{-M}$, where $V_{+M}$ and $V_{-M}$ are taken as the average voltages measured between (0.04T, 0.06T) and (-0.06T, -0.04T) respectively. Figure 3k summaries δV as a function of the bias current *I* for both forward (negative to positive) and backward (positive to negative) magnetic field sweeps. We can see that δV is linearly increasing with increasing current when |*I*|< 10 µA, while the dependence weakens for |*I*|> 10 µA. We note both $V_{+M}$ and $V_{-M}$ are themselves largely linear with *I* (Fig. S3, with small deviation from linearity at larger current more clearly revealed in Fig. 3k). However, their slopes $R_{+M}$= $V_{+M}$/I and $R_{-M}$= $V_{-M}$/I (defining effective 3-terminal resistances for large positive and negative magnetic fields higher than the coercive field of Py) are slightly different (~567.5 Ω and 565.5 Ω respectively, with a difference of δR= $R_{+M}$-$R_{-M}$ ~2 Ω, Fig. S3). This δR can also be directly extracted from the slope of δV vs *I* in its linear portion (Fig. 3k). The linear dependence of the voltage change δV (spin signal) vs bias current *I* confirms that the measurement is in the linear response regime (until *I* becomes too large, when effects such as Joule heating or population of bulk carriers may tend to reduce the spin signal). The interesting observation that the three-terminal resistance depends on the magnetization *M* of the voltage probe (Py) is a manifestation that the underlying channel is spin polarized. The positive sign of δR, meaning that $R_{+M}$ is always larger than $R_{-M}$ (measured with both positive and negative currents), reflects the specific manner that the direction of the spin polarization is locked to that of the current (*I*) and is consistent with the spin helicity of TSS at top surface. The magnitude of δR provides an intrinsic measure (independent of current in the linear response regime) of the ability for a specific sample to generate current-induced spin polarization.

We further studied the temperature dependence of the spin signal in device *A*. Fig. 4a shows the voltage measured by the Py spin detector as a function of the in-plane magnetic field under a DC bias current (*I*) of 10 µA at various temperatures (*T*) ranging from 1.6 to 20K (data appropriately offset vertically for clarity). The hysteretic step-like change in the magnetic field dependent voltage can be observed up to at least ~10K. However, the amplitude of the voltage change for both the positive (Fig. 4a) and negative current biases (Fig. S4) decrease with increasing T and becomes nearly zero at T=20 K. The reduction and disappearance of the measured spin signal of TSS at higher temperatures may be attributed to various factors such as contribution from the bulk states and band bending induced Rashba 2EDG (which has opposite spin helicity from TSS) [32,33], increased scattering (due to phonons) and thermal fluctuations of polarized spins [30], etc.

**Discussion and extraction of current-induced spin polarization.** Following the theoretical analysis by Hong *et al*. for spin potentiometric measurements of current-induced spin polarization in the linear response regime [36], the spin signal or the difference in the ferromagnetic detector voltages $V_M$ and $V_{-M}$ that we measure experimentally (Figs. 2-4) are given by $\delta V = (V_{+M} - V_{-M}) = IR_B P_{FM}(\boldsymbol{p} \cdot \boldsymbol{M_u})$. Here, $I$ is the bias current (noting our positive current and positive magnetization directions are opposite to those in Ref. 36), $1/R_B$ is the "ballistic conductance" of the channel given by $e^2/h$ times the number of modes or conduction channels ($\sim k_F W/\pi$ for each Fermi surface, where $k_F$ is the Fermi wave number, $W$ is the width of channel. Note that although the above expression contains a ballistic conductance ($1/R_B$), used to count the number of propagating modes in the channel, the formula is valid for both ballistic and diffusive regimes as discussed in Ref. 36), $P_{FM}$ is the effective spin polarization (relevant for transport measurements) [26,36] of the FM detector (Py), $\boldsymbol{M_u} = -\boldsymbol{e}_y$ is the unit vector along the positive detector magnetization direction (+$\boldsymbol{M}$), and $\boldsymbol{p}=p\boldsymbol{e}_y$ with the (dimensionless) $p$ being the degree of current-induced channel spin polarization (such that $|pI|$ is the magnitude of the spin-polarized current) and the sign of $p$ determined by the spin helicity (negative for TSS and positive for Rashba-2DEG). Alternatively, the equation can be written as $(V_{+M} - V_{-M}) = |I|R_B P_{FM}(|p|\boldsymbol{s} \cdot \boldsymbol{M_u})$, where $\boldsymbol{s}=\boldsymbol{S}/|\boldsymbol{S}|$ is the unit vector in the direction of the channel spin polarization $\boldsymbol{S}$. Our measured spin signal $\delta V = V_{+M} - V_{-M}$ is qualitatively consistent with what this model predicts for TSS in several ways: 1) the sign of $\delta V$ for a given sign of $I$ is consistent with the TSS (negative sign of $p$, and not consistent with Rashba-induced 2DEG, which would give positive $p$ and the opposite sign of voltage change from our observation); 2) reversing the current $I$, $\delta V$ also reverses; 3) $\delta V$ linearly increases as the current bias increases as shown in Fig. 3k for $|I|<\sim 10$ µA. Furthermore, we can quantitatively extract the spin polarization of the surface sate conduction in our BTS221 sample. We estimate $|p| \approx 0.5$ from the data of Fig. 2 based on W=9 µm, $k_F \approx 0.1$ Å$^{-1}$ (estimated from ARPES measurements on our BTS221 bulk crystals [39] as well as Hall measurements in similar flakes), and assuming $P_{FM}$ (Py) $\approx 0.45$ (ref. 42) and taking the current flowing through the top surface channel as the total bias current flowing through our bulk insulating sample (whose bottom surface, in contact to the substrate, is assumed to be more disordered with higher resistance thus not draw a significant portion of the current; note this assumption would *under*estimate $|p|$). We notice that such a spin polarization is quite large compared to the values (ranging from <0.01 to ~0.2) reported in other recent experiments [26,30,31]. We further note that the amplitude of our spin signal $\delta V$ is larger than that of the metallic $Bi_2Se_3$ samples [26].

So far we have only observed the spin signal (voltage change $\delta V$) in samples with the tunneling barrier ($Al_2O_3$) under the FM detector (Py) and not in samples fabricated without tunneling barriers (Py directly contacting the top surface TI, with otherwise similar device geometry and structures). Such a thin $Al_2O_3$ layer may effectively protect the TI surface from possible damages or other undesirable effects during Py deposition as well as electrical short by the Py contact during the transport measurements. This is consistent with previous experiments using FM spin detectors [44], and also suggests that our measured spin signal is unlikely caused by local Hall effects from Py.

To conclude, we have performed the spin potentiometric measurements using FM voltage probes on TI thin flakes exfoliated from bulk single crystal $Bi_2Te_2Se$ (BTS221, a high-quality bulk-insulating 3D TI), demonstrating the electrical detection of current-induced spin polarization of topological surface state.

The spin signal is manifested as a pronounced difference in the voltage measured at opposite magnetizations of the FM probe, when a DC current is flown through the underlying TI flake. The voltage exhibits a clear hysteretic step-like change as the in-plane magnetic field is swept back and forth to switch the magnetization of the FM probe. We observe a linear dependence of the spin signal on the bias current, and the sign of the spin signal can be reversed by reversing the direction of the current. Our observations provide clear electrical transport evidence for the helical spin current and the spin-momentum locking characteristic of the topological surface states, enabled by the strong spin−orbit interaction and the time-reversal symmetry in 3D TI. Our results may pave the road for further exploration of novel spintronic devices based on TIs and TSS featuring current-induced surface spin polarization.

## Acknowledgements

We acknowledge support for various stages of this work by a joint seed grant from the Birck Nanotechnology Center at Purdue and the Midwestern Institute for Nanoelectronics Discovery (MIND) of Nanoelectronics Research Initiative (NRI), and by DARPA MESO program (Grant N66001-11-1-4107). We are grateful for many helpful discussions with Prof. Supriyo Datta.

**Figure Captions:**

**Figure 1 Schematic of topological surface state and spin potentiometric measurement.** (a) Schematic band structure around the $\Gamma$ point of a topological insulator (TI) near the top surface, showing TSS, bulk valence band (BVB) and bulk conduction band (BCB). Dashed line indicates the Fermi level, $E_f$, and arrows indicate the (in-plane) spin polarization directions of TSS. Inset shows the schematic TSS Fermi surface with its characteristic "left-handed" spin helicity. (b) Schematic 3D device structure used in the potentiometric measurement showing the three-terminal electrical connections as well as the current-induced spin-polarization of the TSS on the top surface. The TI surface defines the x-y plane and the surface normal the z direction. The two outside nonmagnetic (eg., Au) contacts are used to inject bias currents, and the middle ferromagnetic (eg., Py) contact is magnetized by an in-plane magnetic field B (labeled) along its easy axis (y-direction). The middle ferromagnetic contact is a tunneling probe that draws no current. (c-e) Theoretical understanding of the spin potentiometric measurement in the linear response regime probing the current-induced spin polarization due to TSS on the top surface of a 3D TI

for both (d) positive and (e) negative bias currents, based on the spin-dependent electrochemical potentials.

**Figure 2 Electrical detection of the current-induced spin polarization of topological surface state.** (a, b) Voltage measured by the FM (Py) electrode (spin detector) as a function of in-plane magnetic field (along the easy axis of the Py electrode) on a 40 nm–thick flake of $Bi_2Te_2Se$ (BTS221) (device *A*) with bias currents of 10 µA (a), and -10 µA (b). Inset in (a) shows image of the device, with two outer Au electrodes to inject the current and the middle $Py/Al_2O_3$ tunneling probe as detector for the spin potential. The directions of bias current *I*, current-induced spin polarization *s* of TSS, and Py magnetization *M* are labeled by corresponding arrows. Arrows on the data traces indicate magnet field sweep directions. All the measurements are performed at T=1.6 K.

**Figure 3 Current dependence of the measured spin signal.** (a-j) Voltage measured by the Py spin detector on device *A* as a function of in-plane magnetic field at different bias currents of 0.5 µA (a), -0.5 µA (b), 1 µA (c), -1 µA (d), 5 µA (e), -5 µA (f), 20 µA (g), -20 µA (h), 40 µA (i), -40 µA (j). (k) Amplitude of the voltage change $\delta V=V_{+M}-V_{-M}$ as a function of the applied bias current. Here, $V_{+M}$ and $V_{-M}$ are the average voltages measured between (0.04T, 0.06T) and (-0.06T, -0.04T) respectively. The red/black symbols represent the result extracted from the red/black curves (forward/backward magnetic field sweep). Dashed lines are linear fits to the data between -10 µA and 10 µA (in the liner response regime). All the measurements are performed at T=1.6 K.

**Figure 4 Temperature dependence of the measured spin signal.** (a) Voltage (V) measured by the Py spin detector on device *A* as a function of in-plane magnetic field under a DC bias current (*I*) of 10 µA at various temperatures (*T*) ranging from 1.6 to 20K. The five higher-*T* (from 3.6K to 20K) curves are vertically offset by consecutive integer multiples of 35 µV for clarity. The right axis shows the corresponding resistance (R=V/I) as a function of in-plane magnetic field. (b) Amplitude of the voltage change δV (left axis) as a function of temperature with corresponding resistance difference δR=δV/I shown on the right axis.

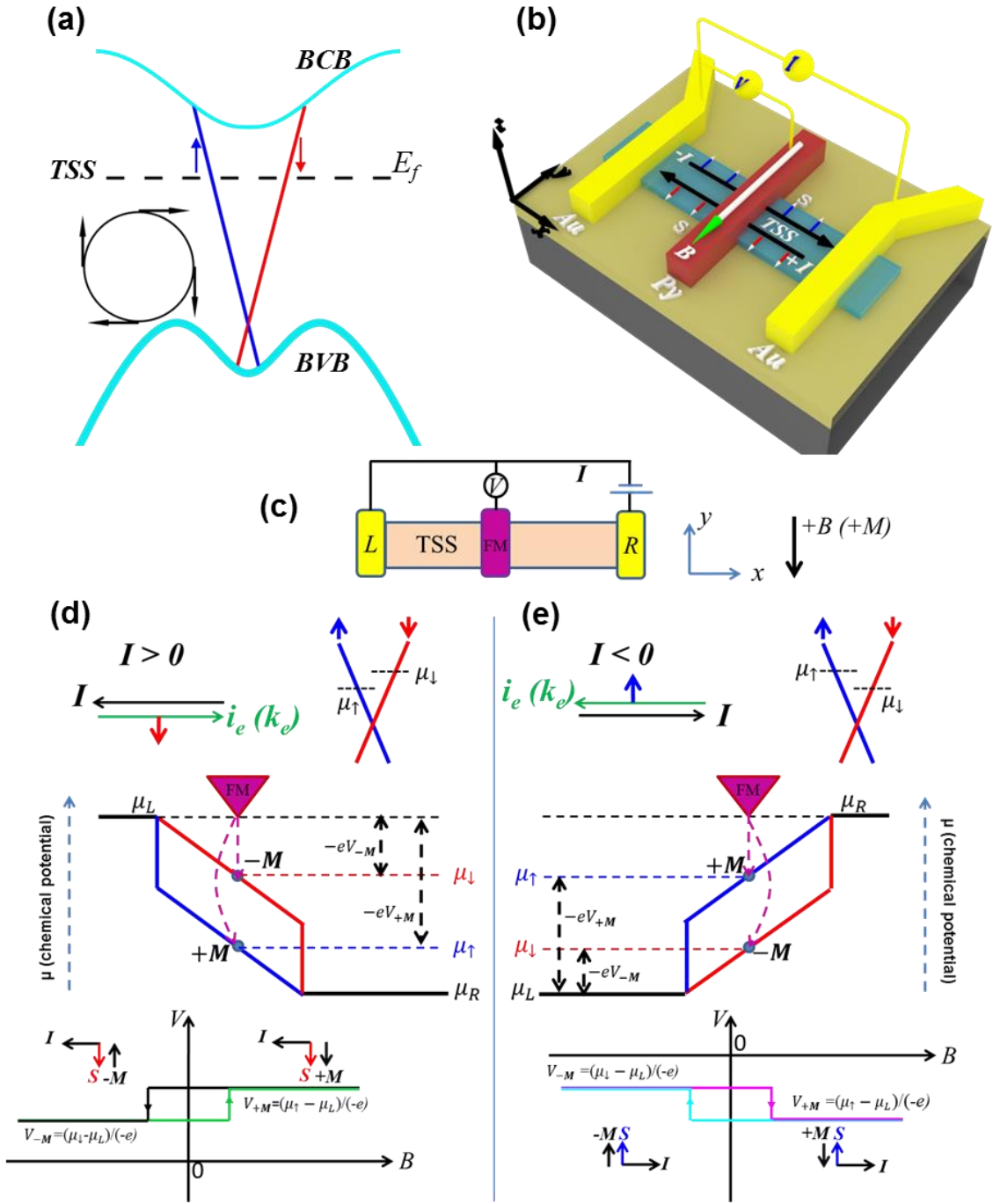

(Figure 1 Tian et al.)

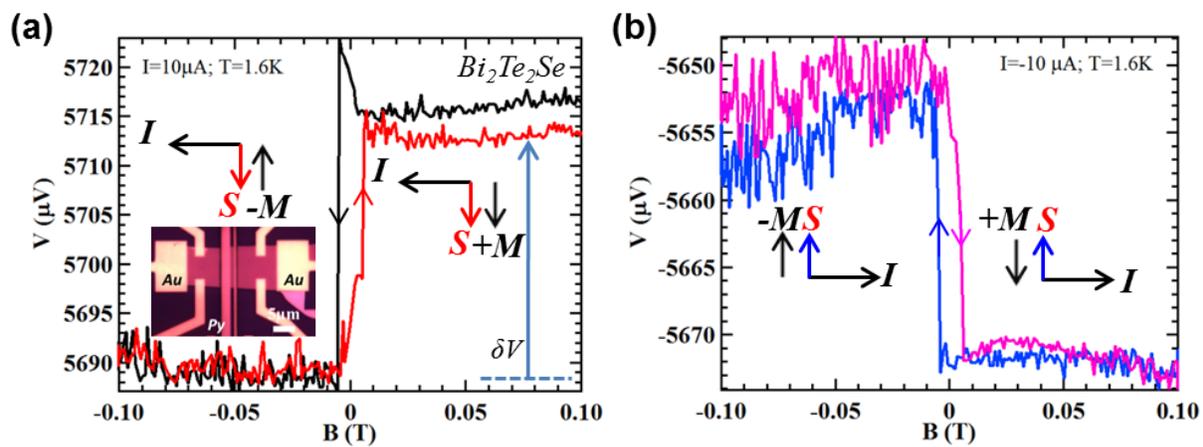

(Figure 2 Tian et al.)

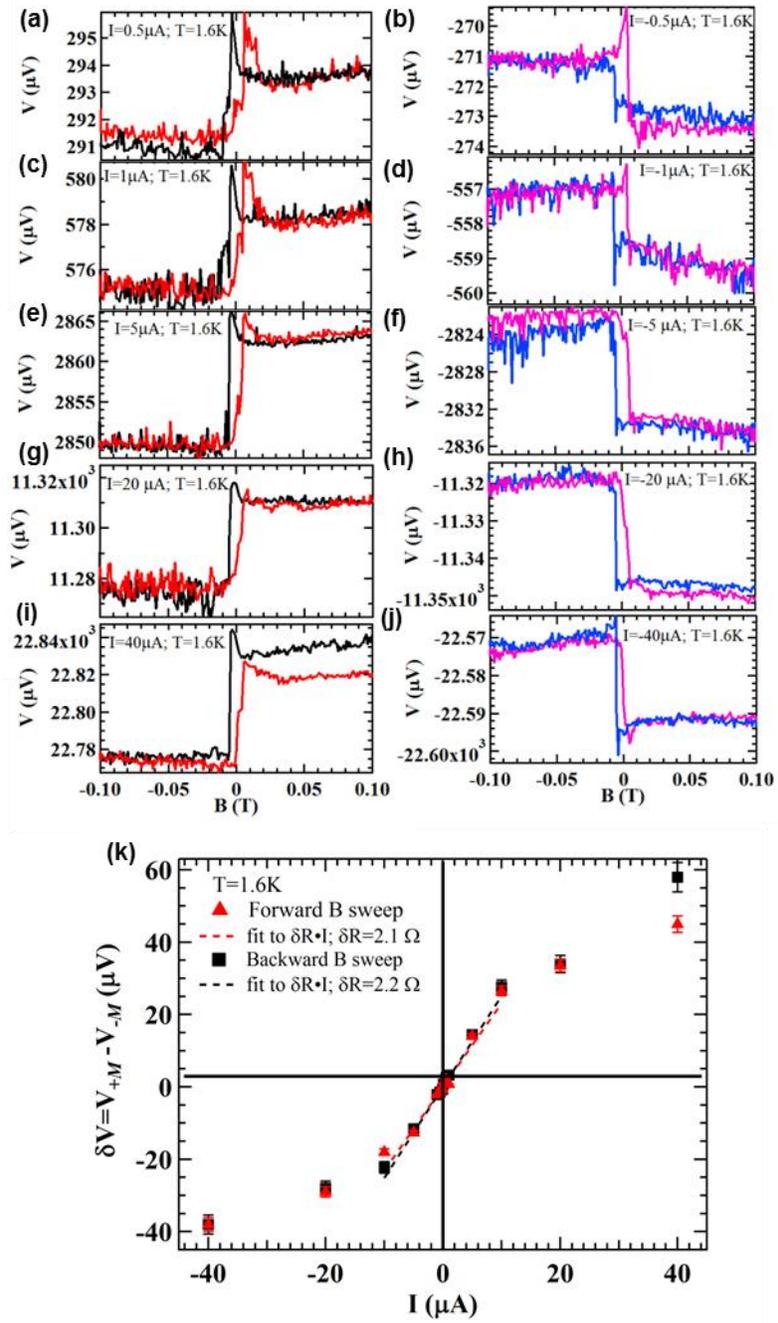

(Figure 3 Tian et al.)

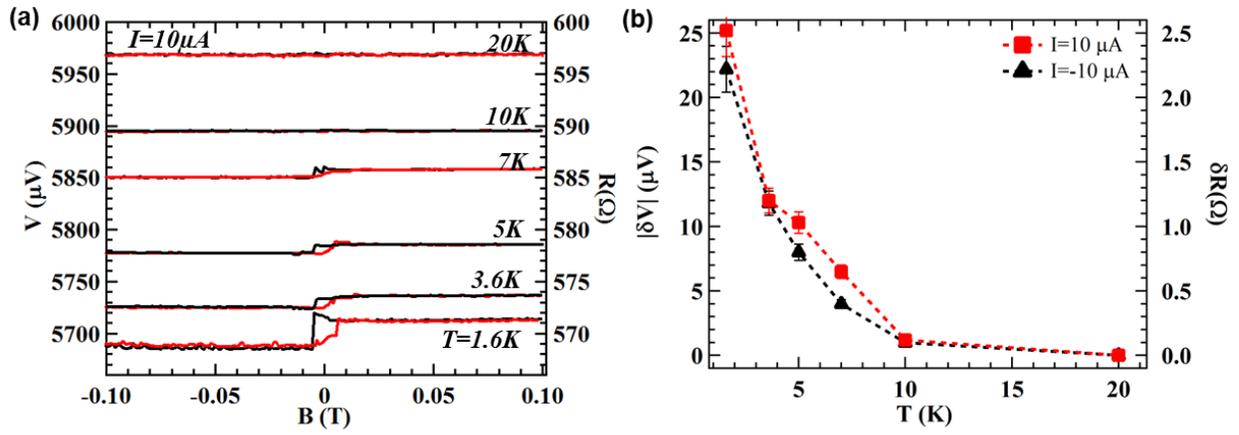

(Figure 4 Tian et al.)

Supplemental Materials

**Electrical injection and detection of spin-polarized currents in topological insulator $Bi_2Te_2Se$**

Jifa Tian[1,2,*], Ireneusz Miotkowski[1], Seokmin Hong[2,3], and Yong P. Chen[1,2,3,*]

1. Department of Physics and Astronomy, Purdue University, West Lafayette, IN 47907, USA

2. Birck Nanotechnology Center, Purdue University, West Lafayette, IN 47907, USA

3. School of Electrical and Computer Engineering, Purdue University, West Lafayette, IN 47907, USA

*Email: tian5@purdue.edu; yongchen@purdue.edu

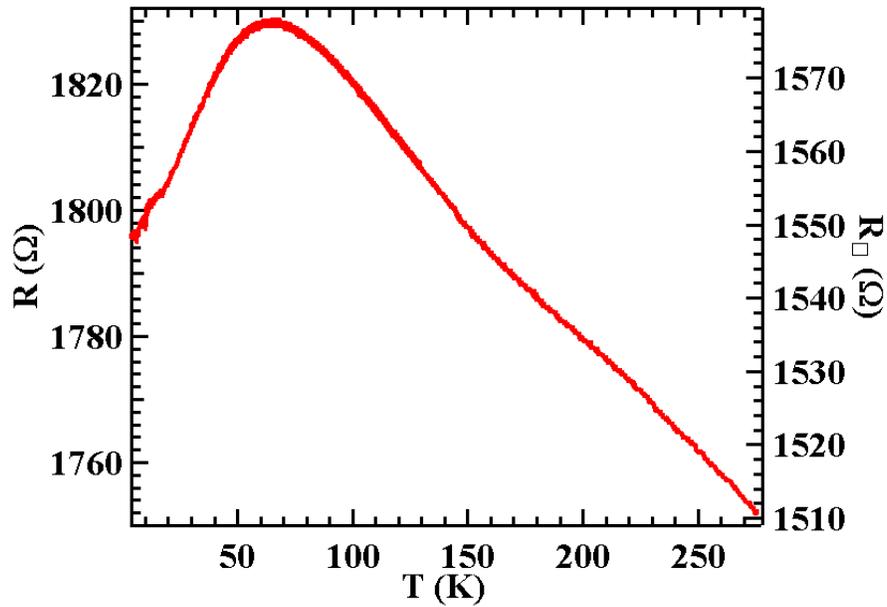

Figure S1 Temperature dependence of the resistance of an exfoliated 39nm-thick BTS221 flake. Right axis shows corresponding sheet resistance. Note such reference devices have only Au contacts and did not go through as many fabrication steps compared to the spin devices in the main text.

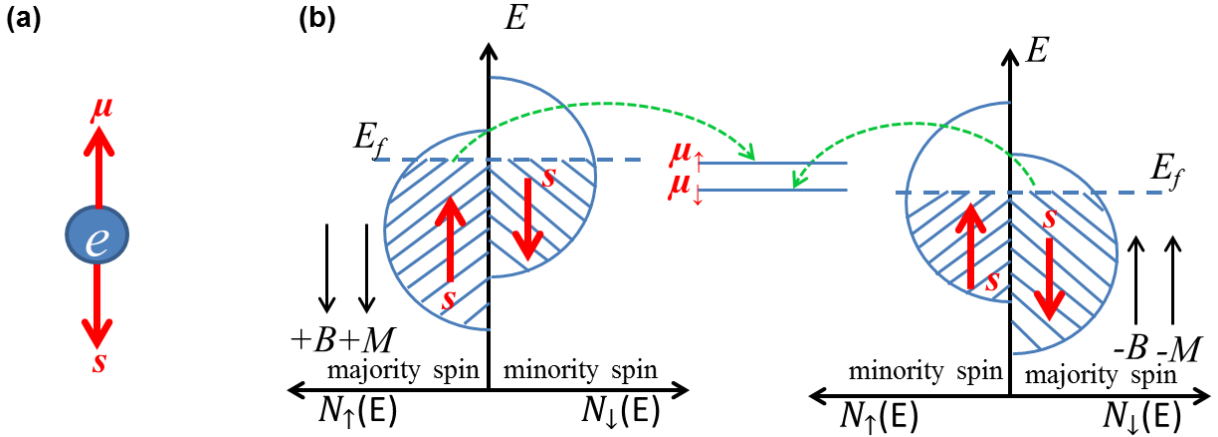

Figure S2 Schematics of magnetization and spin polarization of ferromagnet (FM) and how it measures a spin polarized channel. (a) Schematic representation of the relative orientation of magnetic moment and spin of an electron. The magnetic moment $\boldsymbol{\mu}$ is opposite to its spin $\boldsymbol{s}$ (the direction of the spin angular momentum) due to the negative charge of electron [S1-S3]. (b) Schematic density of states (DOS) diagrams of a ferromagnetic with $+M$ (shown on the left) and $-M$ (right), respectively. The orientation of the majority spin (which determines the magnetization) is opposite to the magnetization direction. A channel with finite "up" spin polarization is depicted in the middle (with more occupation, or higher chemical potential of the "up" spins). When used as a detector (voltage probe) in spin transport (spin potentiometric) measurements, the FM will mainly connect with (and measure the corresponding chemical potential of) the channel spins whose orientation is parallel with that of the FM majority spins [S3-S7]. For example, the FM with down (up) magnetization $+M$ (or $-M$), or equivalently up (down) majority spin orientation, will mainly measure the up (down) spin electrochemical potential in the channel (depicted by the dashed connections).

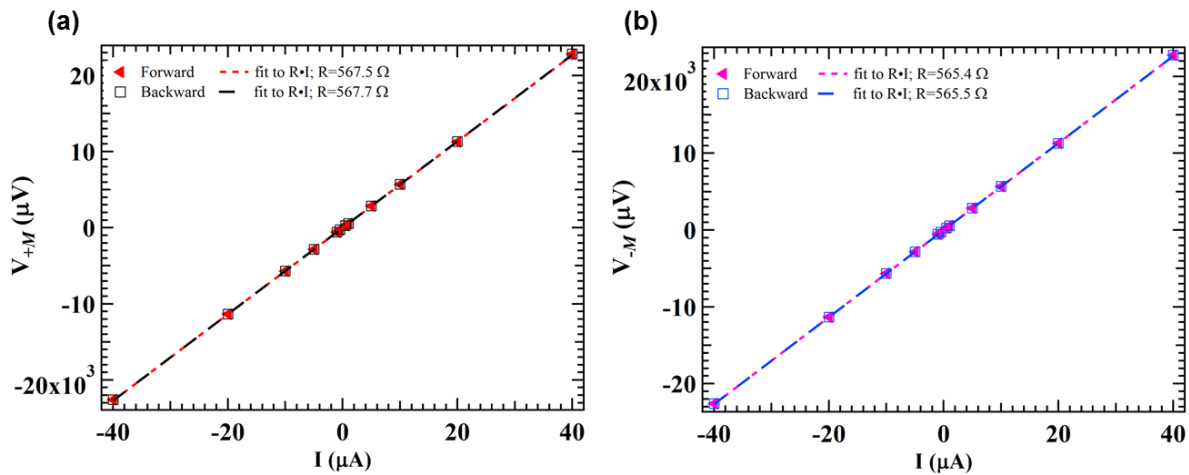

Figure S3 (a) $V_{+M}$ and (b) $V_{-M}$ measured by the Py spin detector on device $A$ as a function of the bias current $I$ for both forward and backward magnetic field sweeps, respectively. Here, $V_{+M}$ and $V_{-M}$ are the average voltages measured between (0.04T, 0.06T) and (-0.06T, -0.04T) as shown in Fig. 3, respectively.

The dashed lines in (a) and (b) are the corresponding linear fittings to R×I. We extract R to be 567.5 Ω (forward) or 567.7 Ω (backward) for $V_{+M}$, and 565.4 Ω (forward) or 565.5 Ω (backward) for $V_{-M}$. The difference between R for +*M* and −*M* is δR~2 Ω. Due to the very thin $Al_2O_3$ used, the IV can still be largely ohmic.

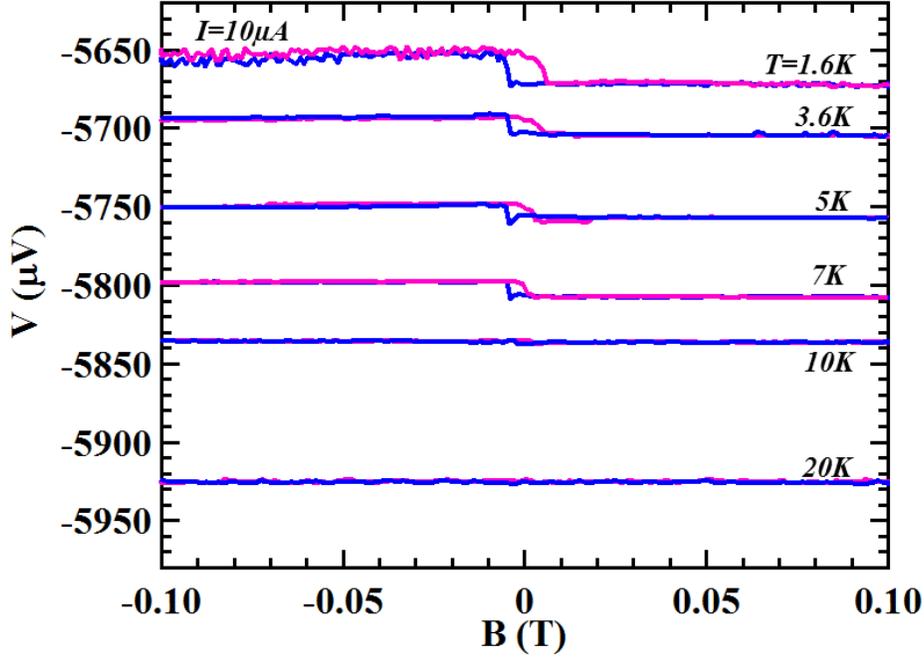

Figure S4 Voltage measured by the Py spin detector on device *A* as a function of in-plane magnetic field under a DC bias current of -10 μA at the same set of temperatures shown in Fig. 4a (from 1.6 to 20K). The five higher-*T* (from 3.6K to 20K) curves are vertically offset by consecutive integer multiples of -35 μV for clarity.

References:

S1 C. Kittel, "Introduction to solid state physics". 8th ed., Wiley, (2004).

S2 B. T. Jonkera, A. T. Hanbickia, D. T. Pierceb, M. D. Stiles, "Spin nomenclature for semiconductors and magnetic metals", J. Mag. Mag. Mater. 2004, 277, 24–28

S3 Handbook of Spin Transport and Magnetism, edited by Tsymbal E. Y.; Zutic, I. CRC Press 2011.

S4 R. Meservey and.P. M. Tedrow, "Spin-polarized electron tunneling", Phys. Rep. 1994, 238, 173-243.

S5 P. R. Hammar and M. Johnson, "Potentiometric measurements of the spin-split subbands in a two-dimensional electron gas", Phys. Rev. B 2000, 61 7207.

S6 M. Johnson, "spin injection and detection in a ferromagnetic metal/2DEG structure", Physica E 2001, 10, 472-477

S7 Y. H. Park, H. C. Jang, H. C. Koo, H.-J. Kim, J. Chang, S. H. Han and H.-J. Choi, "Observation of gate-controlled spin-orbit interaction using a ferromagnetic detector", J. Appl. Phys. 2012, 111, 07C317.